\documentclass[nofootinbib,aps,prd,preprint,superscriptaddress]{revtex4}
\usepackage{epsfig,epstopdf}
\usepackage{amsmath,graphicx}

\newcommand{\n}{\overline{n}}
\newcommand{\hs}{\hat{s}}

\newcommand{\mM}{\mathcal{M}}
\newcommand{\mO}{\mathcal{O}}
\newcommand{\mP}{\mathcal{P}}

\newcommand{\mB}{\mathcal{B}}
\newcommand{\mL}{\mathcal{L}}

\newcommand{\mW}{\mathcal{W}}
\newcommand{\mY}{\mathcal{Y}}

\begin{document}

\baselineskip 3.0ex

\vspace*{18pt}

%%%%%%%%%%%%%%%%%%%%%%%%%%%%%%%%%%%%%%%%%%%%%%%%%%%%%%%%%%%%%%%%%%%%%%
%%%%%%%%%%%%%%%%%%%%%%%%%%%%% Title %%%%%%%%%%%%%%%%%%%%%%%%%%%%%%%%%%
%%%%%%%%%%%%%%%%%%%%%%%%%%%%%%%%%%%%%%%%%%%%%%%%%%%%%%%%%%%%%%%%%%%%%%

\title{Pair Production of Color-Octet Scalars at the LHC}

\def\Duke{Department of Physics, Duke University, Durham NC 27708, USA}
\def\Madrid{Departamento de F«õsica Te«orica II, Universidad Complutense de Madrid, 28040 Madrid, Spain}
\def\CERN{Theory Division, Department of Physics, CERN, CH-1211, Geneva 23, Switzerland\vspace{0.5cm}}

\author{Ahmad Idilbi}\email{idilbi@phy.duke.edu}\affiliation{\Duke}\affiliation{\Madrid}
\author{Chul Kim}\email{chul.kim@cern.ch}\affiliation{\Duke}\affiliation{\CERN}
\author{Thomas Mehen}\email{mehen@phy.duke.edu}
\affiliation{\Duke}

\preprint{CERN-PH-TH/2010-153}

\begin{abstract}

Heavy colored scalar particles, which   exist in many models of new physics, can be pair produced
at the Large Hadron Collider (LHC) via gluon-gluon fusion and possibly form quarkonium-like bound states.
If the scalars are also charged under the electroweak gauge group,  these bound states can then decay into electroweak bosons. This yields a resonant cross section for final
states such as $\gamma \gamma$ that can exceed Standard Model backgrounds.  This paper studies this process in the Manohar-Wise model  of color-octet scalars (COS). Important threshold logarithms and final state Coulomb-like QCD interactions
are resummed using effective field theory.
We compute the resummed cross section for  gluon-gluon fusion to COS pairs at the LHC  as well as the resonant cross section for octetonium decaying to $\gamma \gamma$. The latter cross section exceeds the Standard Model di-photon cross section when the
COS mass is less than 500 (350) GeV for $\sqrt{s} = 14\, (7)$ TeV. Nonobservation of resonances below these energies can significantly
improve existing bounds on COS masses.

\end{abstract}

\maketitle

%%%%%%%%%%%%%%%%%%%%%%%%%%%%%%%%%%%%%%%%%%%%%%%%%%%%%%%%%%%%%%%%%%%%%%

\section{Introduction}

One of the main goals of the Large Hadron Collider (LHC) is to search for new physics (NP) around or above the $1$ TeV scale.
Many new physics models predict heavy scalars carrying a color charge.
Such scalars exist in supersymmetric theories \cite{Plehn:2008ae,Choi:2008ub}, Pati-Salam unification \cite{Povarov:2007nh,Popov:2005wz},
 grand unified theories \cite{Dorsner:2007fy,Perez:2008ry,FileviezPerez:2008ib}, chiral color \cite{Frampton:1987ut}, and top color
 \cite{Hill:1991at}.
 Generically, such particles can introduce unwanted flavor changing neutral currents (FCNC) and the usual expectation is that these
 particles must be quite heavy to avoid experimental  constraints on FCNC. However, these constraints will depend on undetermined
 Yukawa couplings, and if suitable restrictions on these Yukawa couplings are imposed,  the  additional
 scalars can be surprisingly light.  For example, FCNC constraints can be naturally avoided if one imposes Minimal Flavor Violation
 \cite{Chivukula:1987py,D'Ambrosio:2002ex} on the Yukawa couplings of new physics
 to Standard Model (SM) fermions.  Manohar-Wise~\cite{Manohar:2006ga} recently proposed an extended scalar sector with color-octet scalars (COS) that are also  electroweak doublets, the unique representation consistent with MFV.~\footnote{Other representations are possible
 if NP transforms non-trivially under the SM flavor group~\cite{Arnold:2009ay}.} The existence of color-octet scalars
 of this type is weakly constrained by collider phenomenology because the COS couple most strongly to the third generation of quarks. Searches for
 new physics in final states with $b\bar{b}b \bar{b}$ yield a rough constraint of $m_S \gtrsim 200$ GeV, where $m_S$ is the COS mass, assuming that  the COS Yukawa
 couplings to up- and down-type quarks are roughly equal~\cite{Gerbush:2007fe}. Completely model independent constraints are even weaker.
 Ref.~\cite{Burgess:2009wm} concludes that masses of these particles could be as low
 as $\sim$ 100  GeV and still be consistent with precision electroweak fits and collider data.

In a recent paper~\cite{Kim:2008bx}, two of us argued that better constraints on the masses of COS can be obtained in searches
for bound states of the COS. The COS can be pair-produced and  have a strong attractive potential when they are
in a color-singlet state. If the Yukawa couplings of the COS to SM fermions are $O(1)$ or smaller, this state can live long enough to form quarkonium-like bound states called octetonium. These bound states can then decay to pairs of electroweak bosons, e.g., $\gamma \gamma$, $\gamma Z^0$, $W^+ W^-$, etc. Thus the octetonium
would appear as a resonance in these channels which have relatively small SM backgrounds.
The couplings to gluons and electroweak bosons are fixed by gauge symmetry so the only free parameter in the calculation
of the cross section is the COS mass. Ref.~\cite{Kim:2008bx} calculated the production cross section for octetonia via gluon-gluon fusion as
well as decay rates for a number of two-body decays to SM  particles. A back of the envelope
comparison of octetonium production via gluon-gluon fusion followed by decay to $\gamma \gamma$ suggested that the resonant
cross section for this process would exceed the SM contribution for COS masses of 500 GeV or less. Thus, better constraints
on the COS masses than those found in Refs.~\cite{Gerbush:2007fe,Burgess:2009wm} could be obtained from null searches in these channels.

The goal of this paper is to perform a more careful calculation of the process discussed in Ref.~\cite{Kim:2008bx} by incorporating
important QCD corrections that arise in the calculation of pairs of strongly interacting heavy particles near threshold. Many of the
same issues arise in the calculation of $t \bar{t}$, squark-anti-squark, and gluino pair production~\cite{Kiyo:2008bv,Kulesza:2008jb,Kulesza:2009kq,Younkin:2009zn,Beneke:2010gm}.
There are two classes of corrections one needs
to take into account. First, there are (partonic)   threshold logarithms that appear in any production process characterized by a large partonic center-of-mass energy threshold including, e.g.,  Higgs production or Drell-Yan.
In our previous paper \cite{Idilbi:2009cc}, the  resummation of these logarithms for the production of a single COS
was performed using Soft-Collinear Effective Theory (SCET)~\cite{SCET1,SCETf,Bauer:2002nz}.
There we showed that the resummation increased the  normalization of the total cross section by a factor of 2-3
 for a COS with mass in the range 500 GeV-3 TeV. The additional effect that must be taken into account when two heavy colored particles
 are produced is the exchange of Coulomb gluons between the heavy particles in the final state. The exchanges scale as $\alpha_s/v$ where $v$ is the relative velocity of the  heavy particles. In the threshold region, $v \sim \alpha_s$ graphs with Coulomb gluons must be resummed to all orders. The exchange of Coulomb gluons is responsible for the  attractive potential between the COS when they are in a color-singlet state and gives rise to the resonant enhancement of the cross section when the invariant mass of the COS pair is close to that of the octetonium bound state.

The outline of the paper is as follows. In Section II, we perform tree-level matching of the amplitude for $gg\to S^+S^-$ in the Manohar-Wise model onto
SCET and Heavy Scalar Effective Theory (HSET). The resulting  operator couples the COS, which are slowly moving and hence described by HSET fields,
directly to the initial state gluons, which are described by SCET collinear fields. In Section III, we derive a factorization
theorem for $\sigma(pp \to S^+S^-X)$. The cross section factors into a hard part (proportional to the square of the
matching coefficient obtained in Section II), a soft function, and parton distribution functions (PDF's). Exchange of Coulomb gluons
is included in the QCD Coulomb Green's function.  In Section IV, we solve renormalization group equations (RGE) for each of the components in the
factorization theorem. The resummed production cross section up to next-to leading logarithm (NLL) is obtained directly in momentum space using the methods of Ref.~\cite{Becher:2006nr}.
In section V, we extend our results
to the cross section $\sigma(pp \to S^+S^- \to \gamma \gamma)$ and compare with the NLO  SM calculation
of $\sigma(pp \to  \gamma \gamma)$ obtained using the program DIPHOX~\cite{Binoth:1999qq}.
Before continuing, we wish to emphasize the universality of the factorization and resummation. All dependence on the
model of NP is contained in the matching coefficients which enter the hard part of the cross section. The remaining
steps of the calculation are independent of the model of NP. With suitable modification of the hard part, the results
of this paper can be applied to any model of NP that contains COS.

\section{\label{II} Matching $S^+S^-$ production on HSET/SCET Operators}

\begin{figure}[b]
\begin{center}
\epsfig{file=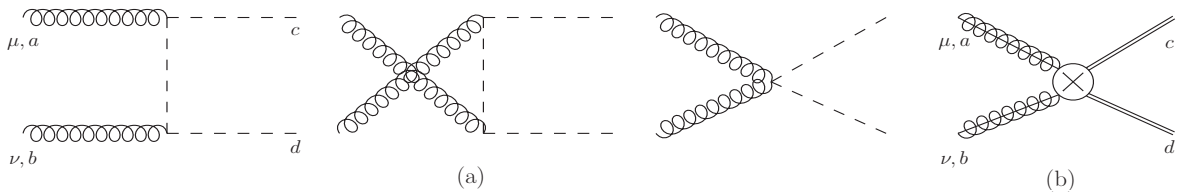, width=16cm, height=2.5cm}
\end{center}
\vspace{-0.7cm}
\caption{\label{fig1} \baselineskip 3.0ex Feynman diagrams for leading order color-octet pair production in full QCD (a) and effective theory (b). }
\end{figure}

%\begin{figure}[t]
%\begin{center}
%\includegraphics[height=6.6cm]{spectatorcontrib}
%\end{center}\vspace{-.5cm}
%\caption{\baselineskip 3.0ex
%Specific examples of DIS processes (a) for a pion and (b) for a proton near the endpoint. In the resonance region $1-x \sim \mO(\eta^2)$, all the final-state quarks are
%$\n$-collinear $(p_c^2 \sim \Lambda^2)$. In the endpoint region $1-x \sim \mO(\eta)$, the spectator quarks in the final state can be either $\n$-hard-collinear
%($p_{hc}^2 \sim Q\Lambda$) or soft ($p_s^2 \sim \Lambda^2$).
%}
%\label{spectatorcontrib}
%\end{figure}

At the LHC, the $gg$ initial state gives the dominant contribution to pair production of COS. The leading order, $O(\alpha_s^2)$, production processes are shown in the diagrams in  Fig.~\ref{fig1}-(a). The couplings come from kinetic terms for the COS,
\begin{equation}
\mL_S = - \frac{1}{2} S^a (D^2)^{ac} S^c - \frac{1}{2} m_S^2 S^a S^a,
\label{SQCD}
\end{equation}
where $D_{\mu}^{ac} = \partial_{\mu} \delta^{ac} + g f^{abc}A_{\mu}^b$. We are interested in calculating the cross section in the threshold region, $\hat{s} \sim (2m_S)^2$, where $\hat{s}$ the momentum squared of incoming partons. In this region, the COS are moving slowly, but the initial state gluons are highly energetic. We integrate out the large scale $m_S$ and the match the diagrams in Fig.~\ref{fig1}-(a) onto operators with
SCET collinear fields describing the initial state gluons and the HSET fields describing the slowly moving COS in the final state.
After this matching, the production of $S^+$ and $S^-$ is accomplished via the operator depicted in Fig.~\ref{fig1}-(b) which is
\begin{equation}
\mL_{\rm{INT}}= \frac{\pi\alpha_s}{2 m_S^3} (f^{kac}f^{kbd} + f^{kad}f^{kbc}) \Bigl(\mY_n \mB_n^{\perp\mu}\Bigr)^a \Bigl(\mY_{\n} \mB_{\n,\mu}^{\perp}\Bigr)^b \Bigl(S_v^{+*}\mY_{v}^{\dagger}\Bigr)^c \Bigl(S_v^{-*}\mY_{v}^{\dagger}\Bigr)^d +\mO(\alpha_s^2),
\label{SCETL1}
\end{equation}
where $n$ and $\n$ are lightcone vectors satisfying $n^2=\n^2=0,~n\cdot \n=2$, and $\mB_{n,\perp}^{a,\mu}$ is a leading $n$-collinear gluon field strength tensor,  defined by $\mB_{n,\perp}^{a,\mu} = i\n^{\rho}g_{\perp}^{\mu\nu} G_{n,\rho\nu}^b \mW_n^{ba} = i\n^{\rho}g_{\perp}^{\mu\nu} \mW_n^{\dagger,ba} G_{n,\rho\nu}^b$. $\mB_{\n,\perp}^{a,\mu}$ is related to $\mB_{n,\perp}^{a,\mu}$
by exchanging $n$ and $\n$. $\mW_n^{ab}$ is a collinear Wilson line in the adjoint representation
\begin{equation}
\mW_n^{ab} (x) = \mathrm{P} \exp \Biggl(ig \int^x_{-\infty} ds \n\cdot A_n^k (s\n^{\mu}) T^k\Biggl)^{ab}.
\label{CW}
\end{equation}

In Eq.~(\ref{SCETL1}), the heavy scalar fields are described by the HSET Lagrangian
\begin{equation}
\label{HSET}
\mL_{\mathrm{HSET}} = S_v^{*a} (v\cdot iD_s)^{ac} S_v^c - \frac{1}{2m_S} S_v^{*a} (D_s^2)^{ac} S_v^c,
\end{equation}
where $v^{\mu}$ is the velocity and $D_s$ is the covariant derivative including only the soft gluon field. The HSET
Lagrangian can be obtained from Eq.~(\ref{SQCD}) by making the substitution
\begin{equation}
\label{Smatching}
S^a (x) = \frac{1}{\sqrt{2m_S}}\Bigl(e^{-im_Sv\cdot x} S_v^a (x) + e^{im_Sv\cdot x} S_v^{*a} (x)\Bigr),
\end{equation}
 dropping all terms in which the large phase does not cancel, and expanding to $O(1/m_S)$.
Note that we have decoupled soft gluons from the collinear and heavy fields by performing field redefinitions,
so the soft Wilson lines
\begin{equation}
\mY_{\mathrm{v}}^{ab} (x) = \mathrm{P} \exp \Biggl(ig \int^x_{-\infty} ds \mathrm{v}\cdot A_s^k (s\mathrm{v}^{\mu}) T^k\Biggl)^{ab},~~~\mathrm{v}^{\mu} = n^{\mu}, \n^{\mu}, v^{\mu}.
\label{SW}
\end{equation}
appear in the operator in Eq.~(\ref{SCETL1})

We find it useful to classify operators by the irreducible representation of color carried by the initial and final states.
The possibilities are enumerated by applying $\bf{8}\otimes \bf{8} = \bf{1} \oplus \bf{8}_S \oplus \bf{8}_A \oplus 10 \oplus \overline{10} \oplus 27$
to the initial and final states, and demanding that total color be conserved. There are eight operators that can contribute to
COS pair production via gluon-gluon fusion, and we will denote them by $(R_i,R_f) =\bf{(1,1)}, (8_S,8_S), (8_S,8_A), (8_A,8_S), (8_A,8_A), (10,10), (\overline{10},\overline{10})$, and $\bf{(27,27)}$~\cite{ Beneke:2009rj}, where $R_{i}$ and $R_f$ denote the  irreducible representations of initial and final states, respectively.  Using this operator basis the interaction Lagrangian in Eq.~(\ref{SCETL1}) is
\begin{equation}
\mL_{\rm{INT}} =\sum_m C_m (\mu) \mO_m (\mu),
\label{SCETL2}
\end{equation}
where each operator is given by
\begin{equation}
\mO_m (\mu) = \frac{1}{2m_S^3} E_{abcd}^{(m)}\Bigl(\mY_n \mB_n^{\perp\mu}\Bigr)^a \Bigl(\mY_{\n} \mB_{\n,\mu}^{\perp}\Bigr)^b \Bigl(S_v^{+*} \mY_{v}^{\dagger} \Bigr)^c \Bigl(S_v^{-*}\mY_{v}^{\dagger}\Bigr)^d.
\label{SCETop}
\end{equation}
Here the color factors $E_{abcd}^{(m)} = E_{abcd}^{(R_i, R_f)}$ are~~\cite{ Beneke:2009rj}
\begin{eqnarray}
E_{abcd}^{(1)} &=& E_{abcd}^{\bf(1,1)} =\frac{1}{8} \delta_{ab}\delta_{cd}, \\
E_{abcd}^{(2)} &=& E_{abcd}^{\bf(8_S,8_S)} = \frac{3}{10\sqrt{2}} D_{ba}^k D_{cd}^k,~~~
E_{abcd}^{(3)} = E_{abcd}^{\bf(8_S,8_A)} = \frac{1}{2\sqrt{10}} D_{ba}^k F_{cd}^k, \nonumber \\
E_{abcd}^{(4)} &=& E_{abcd}^{\bf(8_A,8_S)} = \frac{3}{10\sqrt{2}} F_{ba}^k D_{cd}^k,~~~
E_{abcd}^{(5)} = E_{abcd}^{\bf(8_A,8_A)} = \frac{1}{2\sqrt{10}} F_{ba}^k F_{cd}^k, \nonumber \\
E_{abcd}^{(6/7)} &=& E_{abcd}^{\bf(10,10)/(\overline{10},\overline{10})} = \frac{1}{4\sqrt{10}} \Bigl[\delta_{ac}\delta_{bd}-\delta_{ad}\delta_{bc} -\frac{2}{3}F_{ba}^k F_{cd}^k \pm \Bigl(D_{ac}^k F_{bd}^k + F_{ac}^k D_{bd}^k\Bigr)\Bigl], \nonumber \\
E_{abcd}^{(8)} &=& E_{abcd}^{\bf(27,27)} = \frac{1}{6\sqrt{3}} \Bigl[\delta_{ac}\delta_{bd}+\delta_{ad}\delta_{bc}-\frac{1}{4}\delta_{ab}\delta_{cd} -\frac{6}{5}D_{ba}^k D_{cd}^k \Bigl], \nonumber
\end{eqnarray}
where $D^a_{bc} = d^{abc}, F^a_{bc} = T^a_{bc} = -if^{abc}$, and we set $N_c=3$, where $N_c$ is a number of colors. All the color factors satisfy the orthonormality relation    $E_{abcd}^{(i)}E_{abcd}^{(j)} = \delta^{ij}$.
At tree level the Wilson coefficients in Eq.~(\ref{SCETL2}) are $(C_1,C_2,C_8)= \pi\alpha_s( 6,6\sqrt{2},-6\sqrt{3})$, and $C_3=C_4=C_5=C_6=C_7=0$.
In general, these color factor $E_{abcd}^{(R_1,R_2)}$ are defined to be~~\cite{ Beneke:2009rj}
\begin{equation}
\label{cf}
E_{abcd}^{(R_1,R_2)} = E_{cdab}^{(R_2,R_1)*} = \frac{1}{\sqrt{\mathrm{dim} R_1}} C_{\alpha ab}^{R_1} C_{\alpha cd}^{R_2*},
\end{equation}
where $C_{\alpha ab}^{R}=\langle R, \alpha | ab\rangle$ are the Clebsch-Gordon coefficients between two different color spaces $(R,\alpha)$ and $(a,b)$. The
$E_{abcd}^{(R_1,R_2)}$ vanish unless the two irreducible representations $R_1$ and $R_2$ have the same dimension.

\section{Factorization for COS Pair Production}

For single COS production at threshold~\cite{Idilbi:2009cc}, the only degrees of freedom after integrating out the hard scale
are the collinear initial state partons, soft partons, and a single heavy COS. The interactions of the heavy COS with the initial state collinear
partons via soft gluon exchange are equivalent to a time-like soft Wilson line. The resulting factorization theorem is
a  convolution of two parton distribution functions (PDFs) and soft function multiplied by hard Wilson coefficients. The factorization formula for COS pair production at threshold  is similar, but Coulomb gluon exchanges between two COS in the final state must also be taken into account.
These can be resummed to all orders using the QCD Coulomb Green's function,
\begin{equation}
\label{Green}
G_{R} ({\mathrm{\bf x}, \mathrm{\bf x'}},E)  = \Bigl\langle {\mathrm{\bf x}}\Bigl| \frac{1}{H_R - E}\Bigr| {\mathrm{\bf x'}}\Bigr\rangle
= \sum_n \frac{\psi_n^R (\mathrm{\bf x})\psi_n^{R*} (\mathrm{\bf x'})}{E_n - E-i\epsilon},
\end{equation}
where $H_R$ is the nonrelativistic Hamiltonian including the Coulomb potential in a specific irreducible representation,  $R$, that for a pair of COS must be
${\bf 1, 8_S, 8_A, 10, \overline{10}}$, or ${\bf 27}$.  In the second equality of Eq.~(\ref{Green}), we have written the Coulomb Green's function in terms of the wavefunctions,
$\psi_n^R (\mathrm{\bf x})$,  of eigenstates with energy $E_n$. Our strategy for extracting the dependence of the cross section on the full Coulomb's Green function is to calculate the cross section for energy eigenstates and then use the
second identity in Eq.~(\ref{Green}) to infer the dependence on the full Couloumb's Green's function. We will also take into account the finite width of the  COS by making the replacement $E\to E+i \Gamma_S$.

The cross section in the threshold region for producing COS pairs is given by
\begin{eqnarray}
\label{fac1}
\sigma_t (pp\to SSX) &=& \sum_{R_f={\bf 1}}^{{\bf 27}} \sigma_{R_f} (pp\to SSX) \\
&=&  \sum_{R_f,\alpha} \frac{1}{2s} \sum_{m,k} \sum_X \int \frac{d^3 q}{(2\pi)^3} \frac{1}{2q^0} (2\pi)^4 \delta (P_n +P_{\n} -q - p_X) \Bigl| \mM_{m,k}^{(R_f,\alpha)} \Bigr|^2, \nonumber \\
&=&\sum_{R_f,\alpha} \frac{\pi}{s} \sum_{m,k} \sum_X \delta(q^2 - M_k^2) \Bigl| \mM_{m,k}^{(R_f,\alpha)} \Bigr|^2\Biggl|_{q=P_n+P_{\n}-p_X=p_n+p_{\n}-p_{X_S},} \nonumber
\end{eqnarray}
where $P_{n,\n}^{\mu}$ are the incoming protons' momenta, $p_{n,\n}^{\mu}$ are the momenta of the partons, $M_k$ is a bound state mass, and the matrix elements $\mM_{m,k}^{(R,\alpha)}$ are defined to be $\mM_{m,k}^{(R,\alpha)}=\langle O_k^{(R,\alpha)} X | C_m \mO_m | P_n P_{\n} \rangle$. Here the state $|{\cal O}_k^{(R,\alpha)}\rangle$ is a COS pair in the color state $(R,\alpha)$ and $k$ refers to all other quantum numbers.
The subscript $m$ denotes the $(R_i,R_f)$ quantum numbers of the SCET operators and the summation is nonvanishing when the SCET operator's  $R_f$
is the same as the final state's $R$. For example, if we consider the final state with $R={\bf 8_S}$, $m$ can  be either $m=2~{\bf (8_S,8_S)}$ or $m=5~{\bf (8_A,8_S)}$.

The states $X$ in Eq.~(\ref{fac1}) consist of $n(\n)$-collinear and soft partons, so the final state momentum and the phase space integral can be rewritten as $p_X = p_{X_n} + p_{X_{\n}}+p_{X_S}$ and $\sum_X = \sum_{X_{n}}\sum_{X_{\n}}\sum_{X_{S}}$, respectively. The incoming parton momenta satisfy the relations, $p_{(n,\n)} = P_{(n,\n)} - p_{(X_n,X_{\n})}$.
Then the argument of the delta function in the last equality of Eq.~(\ref{fac1}) becomes
\begin{eqnarray}
\label{arg}
q^2-M_k^2 &=& (p_n+p_{\n}-p_{X_S})^2 - M_k^2 \approx \hs - 2\eta\hs^{1/2} - M_k^2  \\
&=& (\hs^{1/2}+M_k)(\hs^{1/2}-M_k)-2\eta\hs^{1/2} \approx 2\hs^{1/2}(\hs^{1/2}-2m_S-E_k -\eta) \nonumber \\
&=& 2\hs^{1/2}(M-2m_S-E_k) = 2\hs^{1/2}(E-E_k), \nonumber
\end{eqnarray}
where $\eta=p_{X_S}^0$, $M_k = 2m_S + E_k$, and the invariant mass of the COS pair is $M = \hs^{1/2}-\eta = 2m_S+E$.

In order to derive the factorization formula in momentum space, we will insert into $\sigma_{R_f}$
\begin{equation}
\label{iden}
1=\int d\eta dy_1 dy_2 \delta(\eta+i\partial_0)\delta\left(y_1 -\frac{\n\cdot \mP}{\n\cdot P_n}\right)
\delta\left(y_2-\frac{n\cdot \mP}{n\cdot P_{\n}}\right),
\end{equation}
where $\n \cdot \mP(n\cdot \mP)$ is a large label operator acting on $n(\n)$-collinear fields,
and the partial derivative, $i\partial_0$, gives the energy of soft partons.
Using the definition of $\mO_i$ in Eq.~(\ref{SCETop}) and the completeness relation $|X\rangle\langle X| =1$, we write $\sigma_{R_f}$ as
\begin{eqnarray}
\label{fac2}
\sigma_{R_f} (pp\to S^+S^- X) &=& \frac{\pi}{(2m_S^3)^2 s} \sum_{m,k} E_{abcd}^{(m)*}E_{efgh}^{(m)}\int d\eta dy_1 dy_2 \delta(q^2 - M_k^2) |C_m(M, \mu)|^2 \nonumber \\
&\times& \Bigl\langle P_nP_{\n}  \Bigl|
(\mB_{n}^{\perp\mu} \mY_{n}^{\dagger})^a (\mB_{\n\mu}^{\perp} \mY_{\n}^{\dagger})^b
(\mY_v S_{v}^+)^c (\mY_v S_{v}^-)^d \Bigr| O_k^{(R_f,\alpha)} \Bigr\rangle  \\
\label{fac3}
&\times& \Bigl\langle O_k^{(R_f,\alpha)} \Bigl| \delta(\eta+i\partial_0) (\mY_{n} \mB_{n}^{\perp\nu}[y_1] )^e (\mY_{\n} \mB_{\n\nu}^{\perp}[y_2] )^f (S_{v}^{+*} \mY_v^{\dagger})^g  (S_{v}^{-*}\mY_v^{\dagger})^h
\Bigr| P_nP_{\n} \Bigr\rangle,  \nonumber \\
&=& \frac{\pi}{8(m_S^2)^3 (N_c^2-1)^2}  \sum_{m,k} E_{abcd}^{(m)*}E_{efgh}^{(m)} \int d\eta dy_1 dy_2 \hat{s}\delta(q^2 - M_k^2)   \\
&\times&  |C_m(M, \mu)|^2 f_{g/P} (y_1) f_{g/P} (y_2)
\langle 0 | S_v^{+r} S_v^{-s} | O_k^{(R_f,\alpha)} \rangle \langle O_k^{(R_f,\alpha)} | S_v^{+*i} S_v^{-*j} | 0 \rangle \nonumber \\
&\times& \Bigl\langle 0 \Bigl| \mY_n^{\dagger pa} \mY_{\n}^{\dagger qb} \mY_{v}^{cr} \mY_{v}^{ds}
\delta(\eta+i\partial_0) \mY_n^{ep} \mY_{\n}^{fq} \mY_{v}^{\dagger ig} \mY_{v}^{\dagger jh} \Bigr| 0 \Bigr\rangle, \nonumber
\end{eqnarray}
where $\hs = y_1 y_2 s$. In the third line of Eq.~(\ref{fac2}), we introduced the following notation
\begin{equation}
\mB^{\perp\mu,a}_n [y_1] = \Bigl[\delta\Bigl(y_1-\frac{\n\cdot \mP}{\n\cdot P_n}\Bigr)\mB^{\perp\mu,a}_{n}\Bigr],~~~\mB^{\perp\mu,a}_{\n} [y_2] = \Bigl[\delta\Bigl(y_2-\frac{n\cdot \mP}{n\cdot P_{\n}}\Bigr)\mB^{\perp\mu,a}_{\n}\Bigr].
\end{equation}
The parton distribution function (PDF) for the gluon in Eq.~(\ref{fac3}) is defined by
\begin{equation}
\langle P_{n} | \mB_{n}^{\perp\mu a} \mB^{\perp\nu b}_{n} [y] | P_{n} \rangle = g_{\perp}^{\mu\nu} \delta^{ab} \frac{y (\n\cdot P_n)^2}{2(N_c^2-1)} f_{g/P} (y).
\end{equation}
The same equation, with $n$ and $\n$ exchanged, defines for the $\n$-collinear gluon PDF .

In Eq.~(\ref{fac3}), the bound states $|O^{(R,\alpha)}_n\rangle$ are defined in terms of the COS states by
\begin{equation}
\label{BS}
|O_n^{(R,\alpha)}\rangle = C^{R*}_{\alpha ab} \sqrt{2 M_k} \int\frac{d^3 k}{(2\pi)^3}
\tilde{\psi}_n^R ({\rm \bf k}) | S_v^{+a} ({\rm\bf k}) S_v^{-b} ({-\rm \bf k})\rangle,
\end{equation}
where $\tilde{\psi}_n^R ({\rm \bf k})$ are the nonrelativistic wave functions in momentum space. The COS single-particle states annihilated by the HSET field, $S^a_v$, are related to the states annihilated by the field $S^a$ by $|S_v^a \rangle = (1/\sqrt{2m_S})|S^a \rangle$. In case of the final state with two identical particles such as $|S_v^0 S_v^0\rangle$, the right side of Eq.~(\ref{BS}) should be divided by $\sqrt{2}$. Finally the matrix elements for the color-octet scalars in Eq.~(\ref{fac3}) can be written in terms of the wavefunctions at the origin using
\begin{eqnarray}
\label{scalarmat}
\langle 0 | S_v^{+r} S_v^{-s} | O_k^{(R_f,\alpha)} \rangle \langle O_k^{(R_f,\alpha)} | S_v^{+*v} S_v^{-*w} | 0 \rangle &=& 2M_k C_{\alpha rs}^{R_f*} C_{\alpha vw}^{R_f} |\psi_k^{R_f} (0)|^2, \\
&\approx& 2M \sqrt{\mathrm{dim} R} ~E_{rsvw}^{(R_f,R_f)*}|\psi_k^{R_f} (0)|^2. \nonumber
\end{eqnarray}
Here we identified $M\sim M_k$ in the second equality, ignoring $\mO(1/M)$ corrections. Using the relation
\begin{equation}
\label{delta}
\delta(q^2-M_k^2) \approx \frac{1}{2 \sqrt{\hat{s}}}\delta(E_k-E)
= \frac{1}{2M\pi}\mathrm{Im} \frac{1}{E_k-E-i\epsilon},
\end{equation}
 inserting Eqs.~(\ref{Green}), (\ref{arg}), and (\ref{BS}) into Eq.~(\ref{fac3}), and using
\begin{equation}
\label{imgGreen}
\sum_k \delta(q^2-M_k^2) |\psi_k^{R_f} (0)|^2 = \frac{1}{2M\pi} \mathrm{Im} \, G_{R_f} (0,0,E)\, ,
\end{equation}
we see that the cross section is proportional to the imaginary part of the Coulomb Green's function evaluated at ${\bf x} ={\bf x^\prime} ={\bf 0}$.
Because the COS are  unstable, we make the substitution $E\to E+i\Gamma_S$.

Combining Eqs.~(\ref{fac3}), (\ref{scalarmat}), and (\ref{imgGreen}), we then obtain
\begin{eqnarray}
\sigma_{R_f} (pp\to SS X) &=& \frac{1}{8 m_S^6(N_c^2-1)^2} \sum_{R_i} \int d\eta dy_1 dy_2 \hs |C_{R_i,R_f} (M, \mu)|^2 \nonumber \\
\label{fac4}
&\times& \mathrm{Im}\, G_{R_f} (0,0,E+i\Gamma_S) f_{g/P} (y_1) f_{g/P} (y_2) S_{R_i,R_f} (\eta),
\end{eqnarray}
where we have slightly modified our notation by replacing  $\sum_m$ with  $\sum_{R_i}$ and $C_m$ with $C_{R_i,R_f}$. The function
$S_{R_i,R_f}$ is defined by
\begin{eqnarray}
\label{soft1}
S_{R_i,R_f} (\eta) &=& \sqrt{\mathrm{dim} R} ~E_{abcd}^{(R_i,R_f)*}E_{efgh}^{(R_i,R_f)} E_{rsvw}^{(R_f,R_f)*} \\
&\times& \Bigl\langle 0 \Bigl| \mY_n^{\dagger pa} \mY_{\n}^{\dagger qb} \mY_{v}^{cr} \mY_{v}^{ds}
\delta(\eta+i\partial_0) \mY_n^{ep} \mY_{\n}^{fq} \mY_{v}^{\dagger vg} \mY_{v}^{\dagger wh} \Bigr| 0 \Bigr\rangle. \nonumber
\end{eqnarray}

When we introduce the variable $z=M^2/\hs$, which goes to 1 at threshold, the soft momentum $\eta$ can be rewritten as
\begin{equation}
\eta = \hs^{1/2}-M = \hs^{1/2}(1-z^{1/2}) \sim \frac{\hs^{1/2}}{2} (1-z),~~~z\to 1.
\end{equation}
Using the relation $y_1 y_2 = \tau/z$ and replacing $\int d\eta \to -\int dM$, we find that the differential scattering cross section  is
\begin{eqnarray}
\label{defcross}
\frac{d\sigma_{R_f}}{dM} (pp\to S^+S^-X) &=& \sum_{R_i} H_{R_i,R_f} (M,\mu_F) \frac{M}{(2m_S)^6}
~\mathrm{Im} G_{R_f} (0,0,E+i\Gamma_S,\mu_F) \\
&\times& \tau \int^1_{\tau} \frac{dz}{z} \overline{S}_{R_i,R_f} (1-z,\mu_F) F\Bigl(\frac{\tau}{z},\mu_F \Bigr), \nonumber
\end{eqnarray}
where $\mu_F$ is the factorization scale,  $F(\tau/z)$ is a convolution of two PDF's,
\begin{equation}
\label{lumi}
F(x,\mu_F) =\int^1_{x} \frac{dy}{y} f_{g/p} (y,\mu_F) f_{g/p} (x/y,\mu_F),
\end{equation}
and the dimensionless soft functions, $\overline{S}_{R_i,R_f}(1-z)$, are defined to be $\overline{S}_{R_i,R_f}(1-z) =  (\hs^{1/2}/2) S_{R_i,R_f}(\eta)$, so that
 $\overline{S}_{R_i,R_f}(1-z)=\delta(1-z)$ at tree level. Finally, the hard function $H_{R_i,R_f}$ is
\begin{equation}
H_{R_i,R_f} (M,\mu) = 16 \frac{|C_{R_i,R_f}(M, \mu)|^2}{(N_c^2-1)^2}.
\end{equation}
This factorization formula is one of our main results, and can be extended to other processes with different  initial states such as $q\bar{q}$ and $qq$. Note that to obtain the scattering cross section for the production of two identical particles, such as $S^0 S^0$, the cross section should be divided by 2.
If we restrict the sum over bound states to the  ground state of the singlet channel ($R_f ={\bf 1}$ and $k=0$ corresponding to the state with
principal quantum number $n=1$ and $l=0$),  and use the tree-level soft function,
$\overline{S}_{R_i,R_f} = \delta(1-z)$, Eq.~(\ref{defcross}) becomes
\begin{equation}
\sigma_{\bf 1}^{(0)} (pp\to O_+^0) = \frac{64\pi^3 N_c^2 \alpha_s^2}{(N_c^2-1)^2 M^5} |\psi_0^{\bf 1} (0)|^2 \tau F(\tau).
\end{equation}
This reproduces the tree-level cross section for $pp\to O_+^0$ in Ref.~\cite{Kim:2008bx}.

 In the scattering cross section, the hard function, $H$, the soft function, $\overline{S}$, and the Coulomb Green's function in Eq.~(\ref{defcross}) should be evaluated  at renormalization scales labelled   $\mu_H$, $\mu_S$, and $\mu_C$, respectively.  These scales are chosen so that large logarithms are minimized. Large logarithms are resummed by evolving the hard function from $\mu_H$ to $\mu_F$, the soft function from $\mu_S$ to $\mu_F$, and the Green's function $\mu_C$ to $\mu_F$. This is described in the next section.

\section{Resummation and Numerical Results}

In this section we calculate the resummed scattering cross section to NLL accuracy with leading order (LO) Wilson coefficients. This approximation, called NLL+LO, includes all $\mO(1)$ terms  when the large logarithms are counted as an inverse power of $\alpha_s$, so corrections to NLL+LO are $O(\alpha_s)$ suppressed. Next-to-leading order (NLO) Wilson coefficients have not been calculated for COS pair production. Refs.~\cite{Ahrens:2008qu,Ahrens:2008nc} have observed that $\pi^2$-enhanced NLO contributions, which can be inferred from the imaginary parts of anomalous dimensions as we will see below,
are numerically similar in size to the complete NLO  $\alpha_s$ correction. So we will include this contribution in our numerical results, which, based on expectations from previous calculations of Higgs production~\cite{Ahrens:2008nc}, should provide a result numerically consistent with a full
 NLL+NLO calculation.

As shown in Sec.~\ref{II}, the only nonzero LO Wilson coefficients for $gg\to SS$ are $C_1$, $C_2$, and $C_8$, which correspond to the initial and final states ${\bf (1,1)}$, ${\bf (8_S,8_S)}$, and ${\bf (27,27)}$, respectively. Computing anomalous dimensions for the NLL resummation in each of these channels is straightforward,
and the results are
\begin{eqnarray}
\label{g1H}
\gamma_{1H} (\mu) &=& -\Bigl(\frac{\alpha_s}{4\pi} \Gamma_{0}^A +\Bigl(\frac{\alpha_s}{4\pi}\Bigr)^2 \Gamma_{1}^A\Bigr)\ln\frac{\mu^2}{-M^2-i\epsilon}-\frac{\alpha_s}{4\pi}B_{\bf 1}^A, \\
\label{g2H}
\gamma_{2H} (\mu) &=& -\Bigl(\frac{\alpha_s}{4\pi} \Gamma_{0}^A +\Bigl(\frac{\alpha_s}{4\pi}\Bigr)^2 \Gamma_{1}^A\Bigr)\Bigl(\frac{1}{2} \ln\frac{\mu^2}{M^2}+\frac{1}{2} \ln\frac{\mu^2}{-M^2-i\epsilon}\Bigr)-\frac{\alpha_s}{4\pi}B_{\bf 8_S}^A, \\
\label{g8H}
\gamma_{8H} (\mu) &=& -\Bigl(\frac{\alpha_s}{4\pi} \Gamma_{0}^A +\Bigl(\frac{\alpha_s}{4\pi}\Bigr)^2 \Gamma_{1}^A\Bigr)\Bigl(\frac{4}{3} \ln\frac{\mu^2}{M^2}-\frac{1}{3} \ln\frac{\mu^2}{-M^2-i\epsilon}\Bigr)-\frac{\alpha_s}{4\pi}B_{\bf 27}^A,
\end{eqnarray}
where $C_A=N_c$,  $B_{\bf 1}^A =2\beta_0$, $B_{\bf 8_S}^A= 2C_A+2\beta_0$, $B_{\bf 27}^A=16+2\beta_0$, and
$\beta_0$ is the first coefficient of the QCD beta function. Here,  $\Gamma_{0}^A$ and $\Gamma_1^A$ are the first and second coefficients of the cusp anomalous dimension of Wilson lines in the adjoint representation: $\Gamma_0^A=4C_A$ and $\Gamma_1^A=8N_c[(67/18-\pi^2/6)N_c-5n_f/9]$, where $n_f$ is a number of flavors.

From the anomalous dimensions we can infer the form of the double  logarithms in the Wilson coefficients, which are
\begin{eqnarray}
C_{\{1,2,8\}} (\mu) = C_{\{1,2,8\}}^{(0)} \Biggl[1&-&\frac{\alpha_s}{4 \pi} C_A \Bigl(\Bigl\{1,\frac{1}{2},-\frac{1}{3}\Bigr\}\ln^2 \Bigl(\frac{\mu^2}{-M^2-i\epsilon}\Bigr) \nonumber \\
\label{Cexp}
&&+\Bigl\{0,\frac{1}{2},\frac{4}{3}\Bigr\}\ln^2 \Bigl(\frac{\mu^2}{M^2}\Bigr)+\cdots \Bigr)\Biggr].
\end{eqnarray}
These lead to large $\pi^2$-enhanced corrections when evaluated at the scale $\mu =M$,
\begin{eqnarray}
\label{Cpi}
|C_{\{1,2,8\}}(M)|^2 &\sim& |C_{\{1,2,8\}}^{(0)}(M)|^2 \Bigl(1+ \frac{\alpha_s \pi}{2}C_A\Bigl\{1,\frac{1}{2},-\frac{1}{3}\Bigr\}\Bigr) \\
&\sim& |C_{\{1,2,8\}}^{(0)}(M)|^2 \exp \Bigl(\frac{\alpha_s\pi}{2}C_A\Bigl\{1,\frac{1}{2},-\frac{1}{3}\Bigr\}\Bigr). \nonumber
\end{eqnarray}
In the second line we have exponentiated, the $\pi^2$-enhanced terms. This is a consequence of evolving the renormalization scale to a complex value
so as to minimize the logarithms~\cite{Idilbi:2009cc}.
Interestingly,  the Wilson coefficient for the $\bf 27$ channel is suppressed when the $\pi^2$-enhanced contribution is included.

In the resummed cross section, we use the tree level values for the soft functions in Eqs.~(\ref{soft1}) and (\ref{defcross}). However, we need to evolve the soft
functions from the soft scale, $\mu_S$, to the factorization scale, $\mu_F$, and to determine the appropriate $\mu_S$ we will use the one-loop expressions for
the soft-functions:
\begin{eqnarray}
\label{s1loop}
\label{s1}
\overline{S}_{\bf 1,1}(1-z,\mu) &=& \delta(1-z) + \frac{\alpha_s}{\pi} N_c A(1-z,\mu), \\
\overline{S}_{\bf 8_S,8_S}(1-z,\mu) &=& \delta(1-z) + \frac{\alpha_s}{2\pi} N_c \Bigl(2A(1-z,\mu)+ B(1-z,\mu)\Bigr), \\
\overline{S}_{\bf 27,27}(1-z,\mu) &=& \delta(1-z) + \frac{\alpha_s}{2\pi} N_c \Bigl(2A(1-z,\mu)+\frac{8}{3} B(1-z,\mu)\Bigr),
\end{eqnarray}
where the coefficient function $A(1-z,\mu)$ is obtained from soft gluon exchanges between $\mY_{n}$ and $\mY_{\n}$ or $\mY_{n(\n)}$ and $\mY_{v}$, and
$B(1-z,\mu)$ from soft interactions between $\mY_{v}$'s. These coefficient functions are
\begin{eqnarray}
\label{eqA}
A(1-z,\mu) &=& \Bigl(\frac{1}{2} \ln^2\frac{\mu^2}{M^2} -\frac{\pi^2}{4}\Bigr) \delta(1-z) - 2 \ln\frac{\mu^2}{M^2} \frac{1}{(1-z)_+} + 4 \Bigl(\frac{\ln (1-z)}{1-z}\Bigr)_+, \\
\label{eqB}
B(1-z,\mu) &=& \Bigl(\ln\frac{\mu^2}{M^2} + 2\Bigr) \delta(1-z) - \frac{2}{(1-z)_+},
\end{eqnarray}
where the standard plus distributions are used, and ultraviolet (UV) poles have been absorbed into counterterms. Note that these expressions are infrared (IR) finite.
The  general form for the  NLO soft function for the  process  $I_1 I_2 \to F$, where $I_{1}$ and $I_{2}$ denote the color representations of the  initial partons and $F$ denotes  the irreducible representation of the final two heavy particle states, is given by
\begin{equation}
\overline{S}_{\bf (I_1,I_2),F} (1-z,\mu)= \delta(1-z) + \frac{\alpha_s}{2\pi} \Bigl((C_{\bf I_1} + C_{\bf I_2}) A(1-z,\mu)
+ C_{\bf F} B(1-z,\mu)\Bigr),
\end{equation}
where $C_{\bf I_{1,2}}$ and $C_{\bf F}$ are the quadratic Casimir operators for the initial and final representations. Our result agrees with Ref.~\cite{Beneke:2009rj}, where the computation has been performed  in coordinate space.

The Coulomb Green's functions in Eqs.~(\ref{imgGreen}) and (\ref{defcross}) are~\cite{Beneke:1999zr,Kiyo:2008bv}
\begin{equation}
\label{aGreen}
G_{R_f} (0,0,E+i\Gamma_S,\mu) = \frac{\alpha_s (\mu)}{4\pi} C_{R_f} m_S^2 \Bigl[-\frac{1}{2\kappa} + \ln\Bigl(\frac{i \mu}{2m_S \bar{v}}\Bigr) +\frac{1}{2} - \psi(1-\kappa)\Bigr],
\end{equation}
where $\kappa$, $\bar{v}$, and $\psi$ are
\begin{equation}
\kappa = i \frac{C_{R_f} \alpha_s (\mu)}{2\bar{v}},~~~\bar{v}=\sqrt{\frac{E+i\Gamma_S}{m_S}},~~~
\psi (z) = \gamma_E + \frac{d}{dz} \ln\Gamma(z).
\end{equation}
Here $E=M-2m_S$, $\gamma_E$ is the Euler gamma, and $\Gamma(z)$ is the Gamma function.
The  $C_{R_f}$ are the coefficients in the LO Coulomb potential, $V_{C,R_f} (r)= -\alpha_s C_{R_f}/r$,
where $R_f$ refers the representation of the COS pair.  For COS pairs, $C_{\bf 1}=N_c$,
$C_{\bf 8_S}=N_c/2$, and $C_{\bf 27}= -1$, so the COS  pairs in the $\bf 1$ and $\bf 8_S$
 feel an attractive force while COS  pairs in the $\bf 27$  feel a repulsive force.
The appropriate scale for the Coulomb's Green's function is $\mu_C \sim m_S v \sim m_S C_{R_f} \alpha_s (\mu_C)$, where $v$ is the
relative velocity of the COS. In the resummed cross section the Coulomb Green's function needs to be evolved from the scale $\mu_C$ to the
scale $\mu_F$, as indicated in Eq.~(\ref{defcross}). However, the Coulomb Green's function anomalous dimension starts at  $O(\alpha_s^2)$
so its evolution can be neglected in a NLL calculation.

The renormalization group equations (RGEs) for the hard functions, soft functions, and PDF's are solved directly in momentum space
using the methods of Ref.~\cite{Becher:2006nr,Becher:2006mr,Becher:2007ty}. The details of the calculation are very similar to the
calculation of the resummed cross section for single COS production in Ref.~\cite{Idilbi:2009cc} so we simply quote our result for the differential cross section:
\begin{eqnarray}
\frac{d\sigma_{R_f}}{dM} (pp\to S^+S^- X) &=& \frac{M}{(2m_S)^6}
~\mathrm{Im} G_{R_f} (0,0,E+i\Gamma_S,\mu_C)  \nonumber \\
\label{resum}
&&\times~\tau \int^1_z \frac{dz}{z} V_{R_f} (z,M,\mu_F) F(\tau/z, \mu_F),
\end{eqnarray}
where the resummation function, $V_{R_f} (z,M,\mu_F)$, is given by
\begin{equation}
\label{VRf}
V_{R_f}(z,M,\mu_f)=\sum_{R_i} H_{R_i,R_f} (M,\mu_H)  U_{R_i,R_f}(\mu_H,\mu_S,\mu_F)
\tilde{S}_{R_i,R_f}(\partial_{\eta},\mu_s) \frac{z^{-\eta}}{(1-z)^{1-2\eta}}\frac{e^{-2\gamma_E \eta}}{\Gamma (2\eta)}.
\end{equation}
Here
$\tilde S_{R_i,R_f}(\partial_\eta,\mu_S)$ are the Laplace transforms of the soft functions,
and the evolution functions $U_{R_i,R_f}(\mu_H,\mu_S,\mu_F)$ are multiplicative factors that come from
evolving the hard functions from the scale $\mu_F$ to the scale $\mu_H$ and
the soft functions from the scale  $\mu_F$ to the scale $\mu_S$.
Up to NLL accuracy the representations of the initial and final states are the same, so below we will simplify our notation
by replacing  $f_{R_i,R_f}$ with $f_{R_f}$ where $f$ represents either a hard function, soft function, or  evolution function, and suppress the summation over $R_i$ in Eq.~(\ref{VRf}).
For the NLL resummation, the auxiliary parameter $\eta$ is defined to be  $\eta = (\Gamma_0^A/\beta_0) \ln (\alpha_s (\mu_f)/\alpha_s (\mu_s))$
 as in Ref.~\cite{Becher:2006nr}.

\begin{figure}[t]
\begin{center}
\epsfig{file=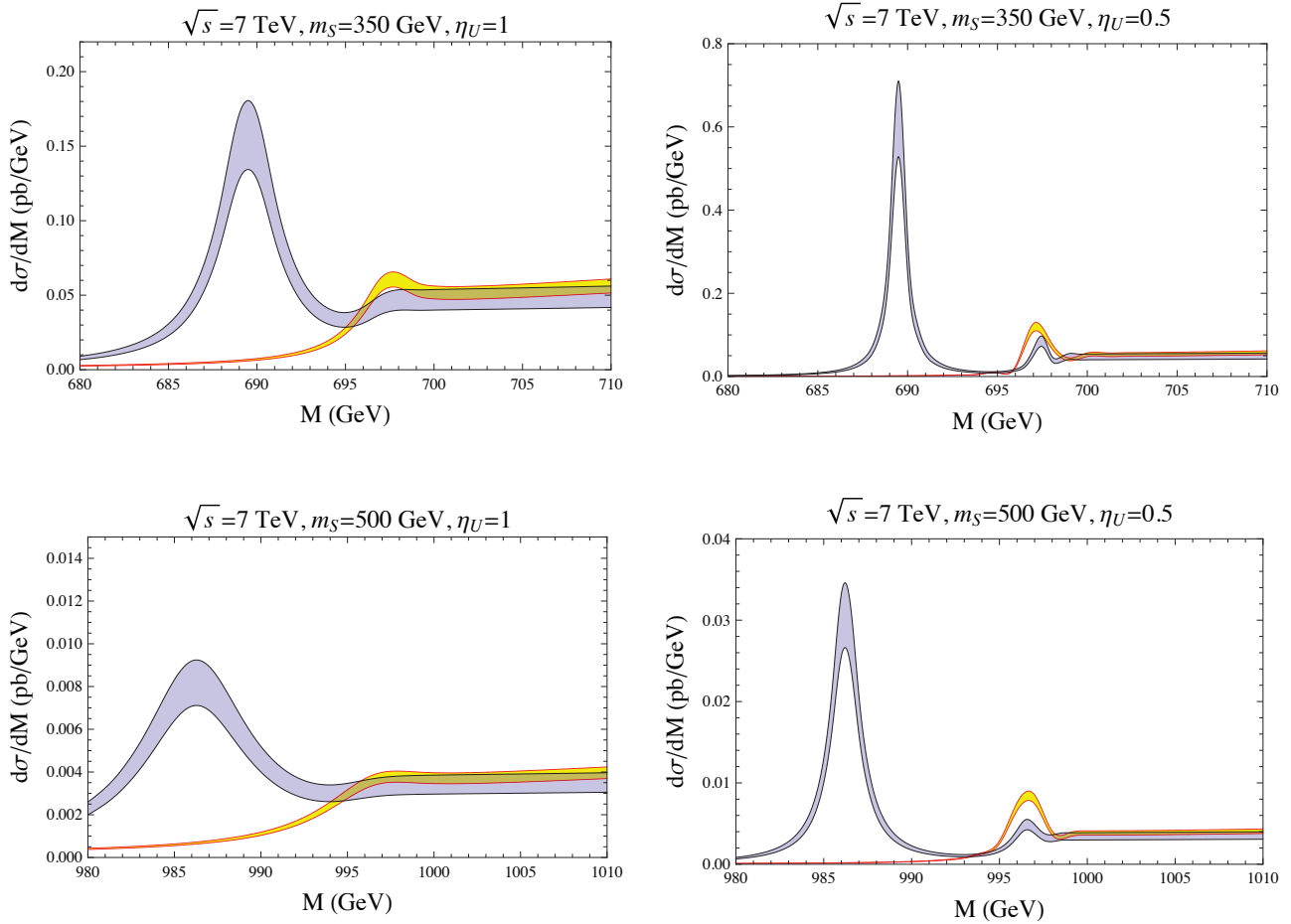, width=17cm}
\end{center}
\vspace{-0.7cm}
\caption{\label{fig2} \baselineskip 3.0ex Mass distribution of the scattering cross section for $pp\to S^+S^-X$ near threshold for $\sqrt{s}=$ 7 TeV. Upper (Lower) filled regions with blue (Yellow) color represent $d\sigma_{\bf 1}/dM~(d\sigma_{\bf 8_S}/dM)$ with the soft scale varied between $\mu_S^{II} \le \mu_S \le \mu_S^{I}$. }
\end{figure}

The NLL expressions for $U_{R_f}(\mu_H,\mu_S,\mu_F)$ are
\begin{equation}
\label{URf}
\ln U_{R_f} (\mu_H,\mu_S,\mu_F) = \ln \Bigl[4 SU_{\mathrm{NLL}}(\mu_H,\mu_S) + \frac{B_{R_f}^A}{\beta_0} \ln \frac{\alpha_s (\mu_S)}{\alpha_s (\mu_H)} + \frac{B_g}{\beta_0} \ln \frac{\alpha_s (\mu_F)}{\alpha_s (\mu_S)}\Bigr],
\end{equation}
where $B_g =2\beta_0$ and $B_{R_f}^A$ are defined in Eqs.~(\ref{g1H}), (\ref{g2H}), and (\ref{g8H}). The function $SU_{\mathrm{NLL}}(\mu_1,\mu_2)$ is
\begin{equation}
SU_{\mathrm{NLL}}(\mu_1,\mu_2) = \frac{\Gamma_0^A}{4\beta_0^2} \Bigl[
\frac{4\pi}{\alpha_s(\mu_1)} \Bigl(1-\frac{1}{r} -\ln r \Bigr)+\Bigl(\frac{\Gamma_1^A}{\Gamma_0^A} -\frac{\beta_1}{\beta_0} \Bigr)(1-r+\ln r) +\frac{\beta_1}{2\beta_0} \ln^2 r \Bigr],
\end{equation}
where $r=\alpha_s (\mu_2)/\alpha_s (\mu_1)$.

\begin{figure}[t]
\begin{center}
\includegraphics[width=17cm]{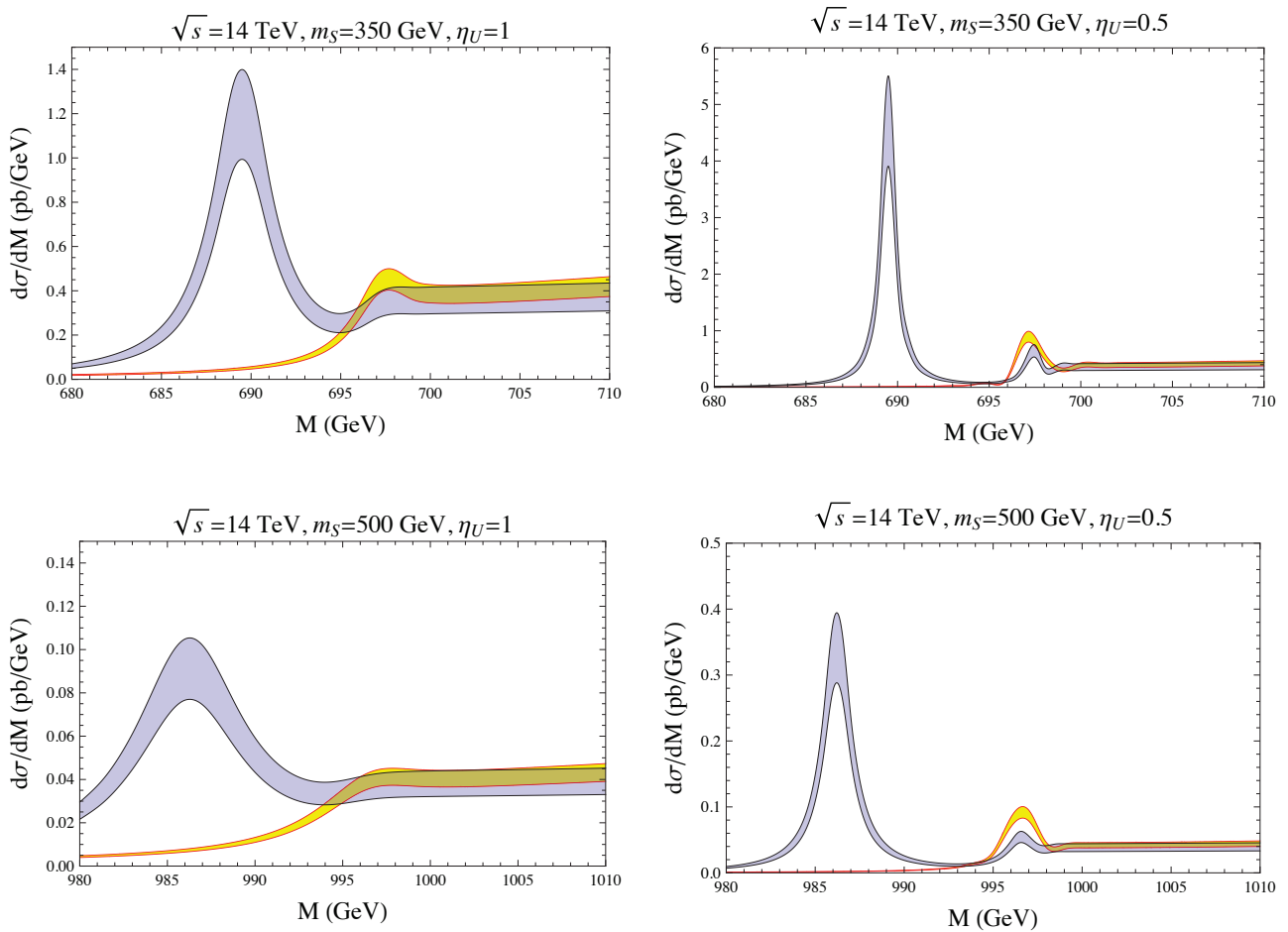}
\end{center}
\vspace{-0.7cm}
\caption{\label{fig3} \baselineskip 3.0ex Mass distribution of the scattering cross section for $pp\to S^+S^-X$ near threshold for $\sqrt{s}=$ 14 TeV. Upper (Lower) filled regions with blue (yellow) color represent $d\sigma_{\bf 1}/dM~(d\sigma_{\bf 8_S}/dM)$ with the soft scale varied between $\mu_S^{II} \le \mu_S \le \mu_S^{I}$. }
\end{figure}

We will choose the hard scale to be $\mu_H=M$, and use the second line of Eq.~(\ref{Cpi}) for the Wilson coefficients in the hard functions so that  the large
$\pi^2$-enhanced contribution is included.
To resum logarithms of $1-z$, the soft scale should be set to $\mu_S= M(1-z)$. However this choice gives divergences in the $z$ integral since the running coupling will cross the Landau pole as $z\to1$. Instead we chose the scale $\mu_S$  so that the higher order corrections to the soft function are perturbatively small.
In order to do this, we define two soft scales, $\mu_S^I$ and $\mu_S^{II}$. The scale $\mu_S^I$ is defined by starting from $\mu_S=\mu_H$ and lowering
$\mu_S$ until the $\mO(\alpha_s)$ correction is less than 15\%. The scale $\mu_S^{II}$ is chosen so that the one-loop correction is minimized. The soft scale
$\mu_S$  is then defined to be the mean of  $\mu_S^I$ and $\mu_S^{II}$~ \cite{Ahrens:2008nc,Becher:2007ty}.

In the  Manohar-Wise model \cite{Manohar:2006ga}, the dominant decay modes for $S^{\pm}$ with masses  greater than $200$ GeV are $S^+ \to t\bar{b}$ $(S^-\to \bar{t}b)$, so the width of the COS is gven by
\begin{equation} \label{width}
\Gamma_S \approx \Gamma (S^{+}\to t\bar{b}) = \frac{|\eta_U|^2|V_{tb}|^2}{16 \pi m_S^3} \Bigl(\frac{m_t}{v}\Bigr)^2 (m_S^2 - m_t^2)^2,
\end{equation}
where $v$ is the vacuum expectation value of Higgs and $\eta_U$ is an unconstrained complex Yukawa coupling. Eq.~(\ref{width})
is used to determine $\Gamma_S$ in the evaluation of the Coulomb Green's function.
When $|\eta_U|$ is smaller than 1, the COS live long enough to form bound states called octetonium~\cite{Kim:2008bx}. Figs.~\ref{fig2} and \ref{fig3}
show the differential cross section as a function of the invariant mass of the COS pair. For the numerical results we employed NLO CTEQ5 PDF set~\cite{Lai:1999wy}.
Fig.~\ref{fig2} shows the  cross section  at $\sqrt{s}=7$ TeV
for $m_S=350$ GeV and 500 GeV, and for two values of the Yukawa coupling, $|\eta_U|=0.5$ and $1.0$. Fig.~\ref{fig3} shows the same for $\sqrt{s}=14$ TeV.
As seen in Figs.~\ref{fig2} and \ref{fig3}, the octetonium
appears as a resonance 10-15 GeV below $2 m_S$ that is clearly visible in the $\bf 1$ channel.
In the $\bf 8_S$ channel, there is a small peak  in the cross section just a few GeV below $2 m_S$.
This peak is so broad for $|\eta_U|=1.0$ that it is barely noticeable, but the peak is visible when the Yuklawa coupling is $|\eta_U|=0.5$.   The scattering cross sections in the ${\bf 27}$ channel
 do not have peaks and  are negligible compared to the ${\bf 1}$ and ${\bf 8_S}$ channels, so we have not included them in Figs.~\ref{fig2} and Fig.~\ref{fig3}.

\section{Scattering Cross Section for $pp\to S^+S^- \to \gamma\gamma$}

Ref.~\cite{Kim:2008bx} argued that the process $pp \to SS \to A B$, where $A B$ represents a pair of SM electroweak gauge bosons, e.g.,
$W^+W^-, Z^0 Z^0, \gamma \gamma,$ or $\gamma Z^0$, are promising channels in which to search for octetonium. Near the vicinity of the
octetonium resonance there is a peak in the cross section which can exceed the SM background for these final states. This is in contrast with final states like $gg$ or
$t\bar{t}$ where we expect the QCD background to greatly exceed any signal from octetonium. In Ref.~\cite{Kim:2008bx}, a simple estimate
for the cross section for  $pp \to S^+S^- \to \gamma \gamma$ in the vicinity of the octetonium resonance was compared with the SM background.
For octetonium with mass $\lesssim 1$ TeV ($m_S \lesssim$ 500 GeV) the cross section at the LHC at $\sqrt{s} = 14$ TeV was found to exceed the SM background for this process. Therefore, searches for $\gamma \gamma$ resonances could either reveal these novel heavy states or provide much better
constraints on the allowed masses of COS, which are currently only constrained to be $\geq 100$ GeV~\cite{Burgess:2009wm}. The point of this
section of the paper is to improve upon the results of Ref.~\cite{Kim:2008bx} by performing a resummed calculation of the invariant mass spectrum
for the photons produced in $pp \to S^+S^- \to \gamma \gamma$ in the vicinity of the octetonium resonance, which is compared with the SM prediction
for the  $\gamma \gamma$ invariant mass distribution.

Below the threshold $2m_S$ (and ignoring the widths of the bound states), the cross section for  $pp\to S^+S^-\to \gamma\gamma$ can be written as a sum over
contributions from individual ${\cal O}^{1}_k$ states,
\begin{equation}
\label{crossgg}
\sigma_{{\bf 1}} (pp\to SS\to \gamma\gamma) = \sum_k \sigma_{{\bf 1},k} (pp\to O_k^{{\bf 1}}X) \frac{\Gamma^{{\bf 1}}_{k} (O_k^{{\bf 1}}\to \gamma\gamma)}{\Gamma^{{\bf 1}}_{k,\rm{tot}}(O_k^{{\bf 1}}\to X)}.
\end{equation}
Note that only color-singlet resonances can decay to the final state $\gamma \gamma$.
Using the factorization formulae in Eq.~(\ref{resum}), integrating over $M$, and writing ${\rm Im} \,G_{R_f={\bf 1}}(0,0,E)$ as a sum of $\delta$-functions times
wavefunctions squared, as in Eq.~(\ref{imgGreen}), we can write $ \sigma_{{\bf 1},k} (pp\to O_k^{{\bf 1}}X)$ as
\begin{eqnarray}
\label{crossk}
\sigma_{{\bf 1},k} (pp\to O_k^{{\bf 1}}X)&=& \frac{2\pi}{(2m_S)^6} \int dM H_{{\bf 1}} (M,\mu) \tau M^2 \\
&&\times \delta(M^2-M_k^2)
|\psi_k ^{{\bf 1}} (0,\mu)|^2
\int^1_{\tau} \frac{dz}{z} \overline{S}_{{\bf 1}} (1-z,\mu) F\Bigl(\frac{\tau}{z},\mu \Bigr). \nonumber
\end{eqnarray}
The LO decay rates for $O_k^{{\bf 1}} \to \gamma\gamma$ are~\cite{Kim:2008bx}
\begin{equation}
\label{decg}
\Gamma^{{\bf 1}}_{k} (O_k^{{\bf 1}}\to \gamma\gamma) =
%\frac{1}{M^2} |\psi_k^{R_f} (0)|^2
%\Bigl[\prod_{j=1,2} \int\frac{d^3p_j}{(2\pi)^3}\frac{1}{2p_j^0} (2\pi)^4\delta(p-p_1-p_2)\Bigr] |\mM_{\gamma\gamma}^{(R_f)} |^2 \nonumber \\
\frac{64\pi \alpha^2}{M^2} |\psi_k^{{\bf 1}} (0)|^2.
\end{equation}
%where the matrix elements $\mM_{\gamma\gamma}^{(R_f)}$ are defined as $\langle \gamma\gamma | H C^{R_f*}_{\alpha ab} | S^{+a}S^{-b} \rangle$.
Finally,  we must allow for a finite width for each of the bound states, ${O}^{{\bf 1}}_k$. We do this by replacing $\delta(M^2-M_k^2)$
with the Breit-Wigner $(M_k \Gamma_{k,\rm{tot}}^{{\bf 1}}/\pi)/((M^2-M_k^2)^2+M_k^2 (\Gamma_{k,\rm{tot}}^{{\bf 1}})^2)$. Then
we can simplify  Eq.~(\ref{crossgg}) with the substitution
\begin{eqnarray}
\sum_{k=0} \delta(q^2-M_k^2) |\psi_k ^{{\bf 1}} (0)|^4 &\to& \sum_{k=0} \frac{\Gamma_{k,\rm{tot}}^{R_f}}{4M\pi}
\frac{|\psi_k ^{{\bf 1}} (0)|^2}{E-E_k+i\Gamma_{k,\rm{tot}}^{{\bf 1}}/2}
\frac{|\psi_k ^{{\bf 1}} (0)|^2}{E-E_k-i\Gamma_{k,\rm{tot}}^{{\bf 1}}/2}  \nonumber \\
&\stackrel{E\sim E_0}{\approx}& \frac{\Gamma_{0,\rm{tot}}^{{\bf 1}}}{4M\pi} \sum_{k=0}
\frac{|\psi_k ^{{\bf 1}} (0)|^2}{E-E_k+i\Gamma_{0,\rm{tot}}^{{\bf 1}}/2}
\frac{|\psi_k ^{{\bf 1}} (0)|^2}{E-E_k-i\Gamma_{0,\rm{tot}}^{{\bf 1}}/2} \nonumber\\
\label{decGreen}
&\approx& \frac{\Gamma_{0,\rm{tot}}^{{\bf 1}}}{4M\pi} |G_{{\bf 1}} (0,0,E+i\Gamma_{0,\rm{tot}}^{{\bf 1}}/2)|^2.
\end{eqnarray}
In the second line we replaced $\Gamma^{{\bf 1}}_{k,{\rm tot}}$ with $\Gamma^{{\bf 1}}_{0,{\rm tot}}$ so we could write the final result
in terms of the Green's function. The corrections to this approximation are not important near the resonance of interest, and small except near the other poles
of the Green's function which should not be important for our calculation.
Then we combine Eqs.~(\ref{crossk}), (\ref{decg}), and (\ref{decGreen}) to obtain
\begin{eqnarray}
\sigma_{{\bf 1}} (pp\to S^+S^- \to \gamma\gamma) &=& \frac{32\pi \alpha^2}{(2m_S)^6} \int \frac{dM}{M} H_{{\bf 1}} (M,\mu) |G_{{\bf 1}} (0,0,E+i\Gamma_{0,\rm{tot}}^{R_f}/2)|^2 \nonumber \\
\label{crossggf}
&&\times~\tau \int^1_{\tau} \frac{dz}{z} \overline{S}_{{\bf 1}} (1-z,\mu) F\Bigl(\frac{\tau}{z},\mu \Bigr).
\end{eqnarray}

\begin{figure}[t]
\begin{center}
\begin{tabular}{cc}
\epsfig{file=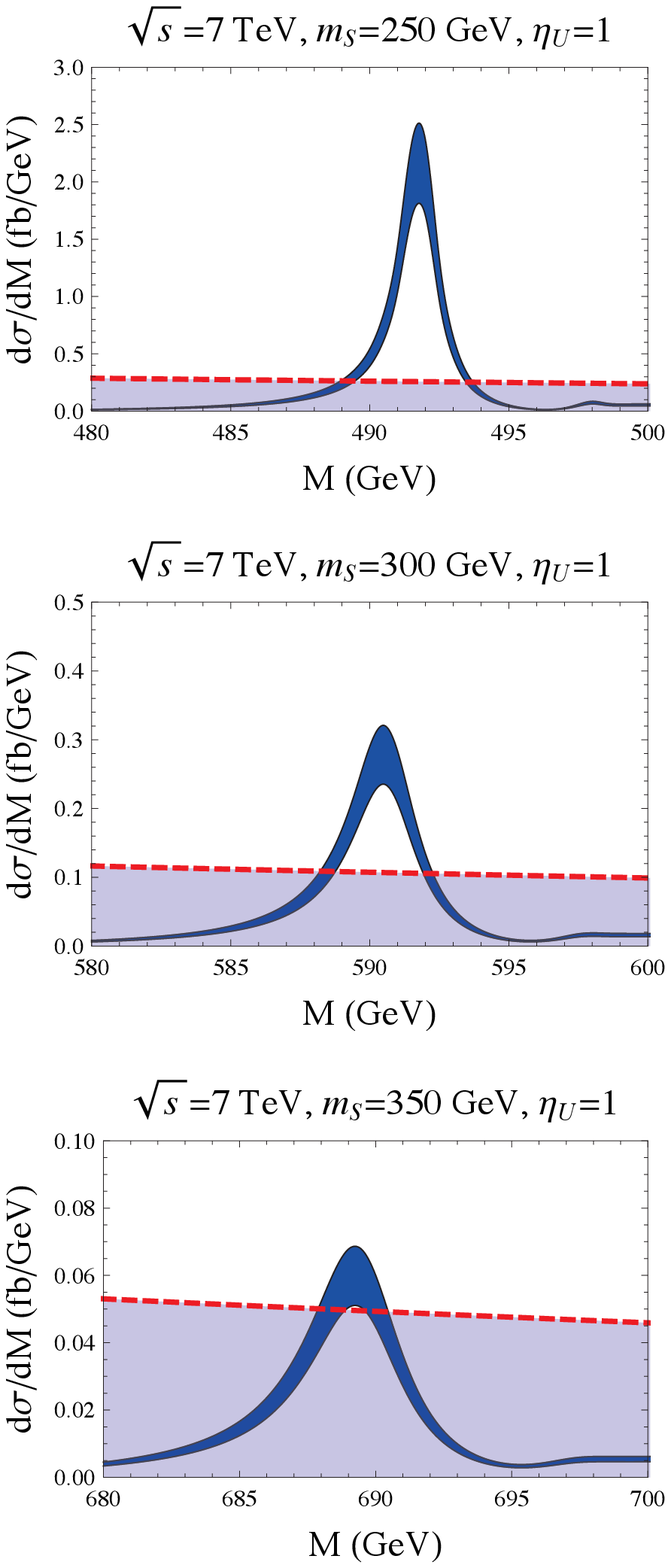, width=8cm} &
\epsfig{file=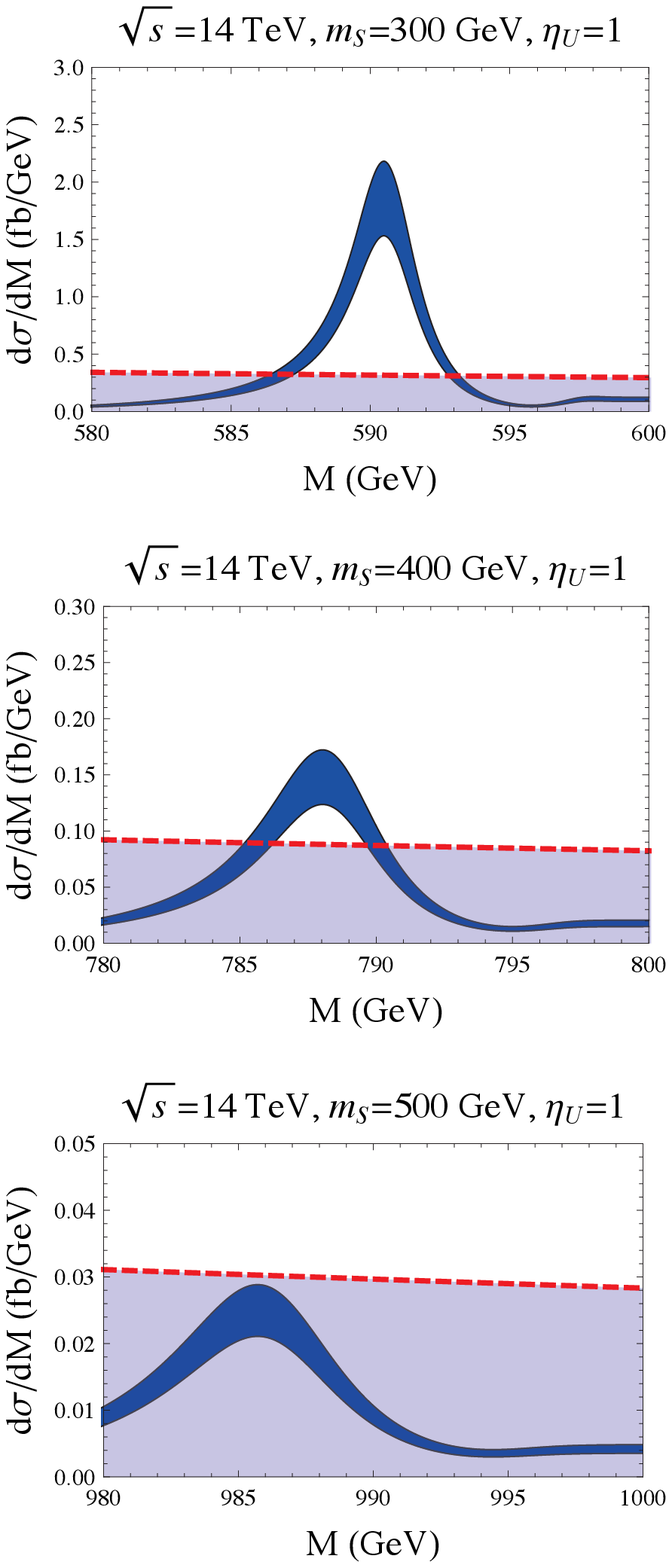, width=8cm}\\
(a) & (b)
\end{tabular}
\end{center}
\vspace{-0.7cm}
\caption{\label{fig4} \baselineskip 3.0ex Mass distribution of the scattering cross section $d\sigma_{\bf 1}/dM (pp\to S^+S^- \to \gamma\gamma)$ near the resonance 
$O_+^0$ ($=O_0^{\bf 1}$) versus the Standard Model background $d\sigma_{\mathrm{SM}}/dM (pp\to \gamma\gamma)$ for (a) $\sqrt{s} = 7~\rm{TeV}$ and (b)
$\sqrt{s} = 14~\rm{TeV}$. }
\end{figure}

Using Eq.~(\ref{crossggf}), we compare the cross section for $pp \to S^+S^- \to \gamma \gamma$ to the SM background $pp\to \gamma \gamma$. The cross section is computed in the vicinity of the $O_+^0(=O_0^{\bf 1})$ resonance in the
Manohar-Wise model. In this model, the width of $O_+^0$ depends on a scalar coupling, $\lambda_1$, which appears in the coupling of a COS pair to the SM Higgs boson~\cite{Manohar:2006ga}. We have set this parameter to $\lambda_1 =1$.
Explicit expressions for the  decay rates for $O_+^0 \to gg,~t\bar{t},~W^+W^-,~Z^0Z^0,~\gamma\gamma$, and $hh$  can be found in Ref.~\cite{Kim:2008bx},
and these have been used to calculate the total width,  $\Gamma_{0,\rm{tot}}^{\bf 1}$.
In Fig.~\ref{fig4}, we compare the $\gamma \gamma$ invariant mass distribution near the peak of the resonance $O_+^0$ with SM backgrounds for $pp\to \gamma\gamma$.
%[[ CHUL: THE FOLLOWING NEEDS TO BE REWRITTEN DESCRIBING THE NEW CALCULATION OF SM BACKGROUND. IT IS NOW NLO AND WE USED THE PROGRAM DIPHOX WHICH NEEDS TO BE CITED.
The SM background cross section, $d\sigma_{\rm{SM}}/dM$, is the sum of NLO calculations of $q\bar{q} \to \gamma\gamma$ and $gg \to \gamma\gamma$ with 
a rapidity  cuts of $|\eta_{1,2}| < 2.4$. The $K$-factor has been computed using the program DIPHOX~\cite{Binoth:1999qq}.
The cross section is computed for $\sqrt{s} = 7$ TeV with  COS masses of 250, 300, and 350 GeV, and
for $\sqrt{s} = 14$ TeV with  COS masses of 300, 400, and 500 GeV. We see that the resonant cross section exceeds the SM contribution
when $m_S \leq 500 \, (350)$ GeV for $\sqrt{s} = 14 \, (7)$ TeV, confirming the conclusions of Ref.~\cite{Kim:2008bx}.
We have used $\eta_U= 1$ in our calculation, for smaller $\eta_U$ the resonance peak is more narrow and visible. Note that the octetonium, ${O}^0_R$,  which is composed of a pair of electrically neutral COS, is significantly narrower than ${O}^0_+$ when $|\eta_U|=1.0$ and $m_S \leq 700$ GeV. Therefore,
 this should appear as a narrower resonance in channels into which it can decay, such as $W^+W^-$ and $Z^0 Z^0$. It would interesting to extend the results
 of this paper to other final states with electroweak bosons.

\section{Conclusions}
The LHC will explore physics beyond the TeV scale. One possibility for new physics that may be discovered at the LHC is the existence of heavy color-octet scalars (COS). In this work we have extended our previous analysis of single COS production~\cite{Idilbi:2009cc} and considered the production cross section of two COS which bind together through Coulomb interactions to form a bound state called octetonium~\cite{Kim:2008bx}. This bound state can decay into two photons, providing a resonant signal above the SM di-photon production cross section. 
 We established a factorization theorem for this production process using SCET and HSET, then performed a next-to-leading logarithmic partonic threshold resummation directly in momentum space. Our factorized cross section is independent of the specifics of the underlying NP theory responsible for the production of COS.
 In this paper, we focused on the Manohar-Wise model of COS, but the calculation can be easily extended 
to pair production of heavy colored particles in other models, e.g. stoponium in supersymmetry~\cite{Martin:2008sv,Martin:2009dj,Kats:2009bv} or 
pairs of Kaluza-Klein excitations of quarks and gluons in models of extra dimensions,
by a suitable modification of the
  matching coefficient at the high scale, $2 M_S$. We find that the resonant cross section exceeds the SM contribution
when $m_S \leq 500 \, (350)$ GeV for $\sqrt{s} = 14 \, (7)$ TeV. Searches for di-photon resonances at the LHC will either discover COS  
resonances or greatly improve existing bounds on COS masses.

%%%%%%%%%%%%%%%%%%%%%%%%%%%%%%%%%%%%%%%%%%%%%%%%%%%%%%%%%%%%%%%%%%%%%%
%%%%%%%%%%%%%%%%%%%%%%%%%%%% Acknowledgments %%%%%%%%%%%%%%%%%%%%%%%%%
%%%%%%%%%%%%%%%%%%%%%%%%%%%%%%%%%%%%%%%%%%%%%%%%%%%%%%%%%%%%%%%%%%%%%%

\acknowledgments

This work was supported in part by the U.S. Department of Energy under
grant numbers DE-FG02-05ER41368 and DE-FG02-05ER41376.
C.~Kim is supported by the Korean-CERN fellowship.

%%%%%%%%%%%%%%%%%%%%%%%%%%%%%%%%%%%%%%%%%%%%%%%%%%%%%%%%%%%%%%%%%%%%%%
%%%%%%%%%%%%%%%%%%%%%%%%%%%%% Bibliography %%%%%%%%%%%%%%%%%%%%%%%%%%%
%%%%%%%%%%%%%%%%%%%%%%%%%%%%%%%%%%%%%%%%%%%%%%%%%%%%%%%%%%%%%%%%%%%%%%

%%%%%%%%%%%%%%%%%%%%%%%%%%%%%%%%%%%%%%%%%%%%%%%%%%%%%%%%%%%%%%%%%%%%%%

\end{document}